\documentclass[preprint, raferee]{aastex}
\usepackage{epsfig,amsmath,amsfonts,amssymb,graphicx,subfigure,enumitem,color}
\usepackage{amssymb}
\usepackage{amsmath}
\usepackage[varg]{txfonts}
\usepackage{natbib}
\usepackage{url}             
\usepackage{graphicx}
\usepackage{float}
\newcommand{\beq}{\begin{equation}}
\newcommand{\eeq}{\end{equation}}
\newcommand{\beqa}{\begin{eqnarray}}
\newcommand{\eeqa}{\end{eqnarray}}
\newcommand{\beqann}{\begin{eqnarray*}}
\newcommand{\eeqann}{\end{eqnarray*}}
\bibpunct{(}{)}{;}{a}{}{,} 

\usepackage{textcomp}
\shorttitle{MRT Unstable Eruptive Prominence}
\shortauthors{Mishra et al.}
\begin{document}
\title{The Evolution of Magnetic Rayleigh-Taylor Unstable Plumes and Hybrid KH-RT Instability into A Loop-like Eruptive Prominence}
\author{Sudheer K.~Mishra}
\author{A.K. Srivastava}
\affil{Department of Physics, Indian Institute of Technology (BHU), Varanasi-221005, India.}
\begin{abstract}
{MRT unstable plumes  are observed in a loop-like eruptive prominence using SDO/AIA observations. The small-scale cavities are developed within the prominence, where perturbations trigger dark plumes (P1 \& P2) propagating with the speed of 35-46 km s$^{-1}$ . The self-similar plume formation shows initially the growth of linear MRT unstable plume (P1), and thereafter the evolution of non-linear single mode MRT unstable second plume (P2). The Differential Emission Measure (DEM) analysis shows that plumes are less denser and  hotter than the prominence. We have estimated the observational growth rate for both plumes as 1.32$\pm$0.29$\times$ 10$^{-3}$ s$^{-1}$  and  1.48$\pm$0.29$\times$ 10$^{-3}$ s$^{-1}$  respectively, which are comparable to the estimated theoretical growth rate (1.95$\times$ 10$^{-3}$ s$^{-1}$) . The nonlinear phase of an MRT unstable plume (P2) may collapse via Kelvin-Helmholtz vortex formation in the downfalling  plasma. Later, a plasma thread has been evident into the rising segment of this prominence. It may be associated with the tangled field and Rayleigh-Taylor instability. The tangled field initiates shearing at the prominence-cavity boundary. Due to this shear motion, the plasma downfall is occurred at the right part of the prominence-cavity boundary.  It triggers the characteristic KH unstable vortices and MRT unstable plasma bubbles propagating at different speeds and merging with each other. The shear motion and lateral plasma downfall may initiate  hybrid KH-RT instability there.}
\end{abstract}

\section{Introduction}\label{sec:intro}
Solar prominences are the cloud of cool and dense plasma materials, which are suspended over the less denser and hot coronal plasma with the support of the magnetic field. The various classifications have been made for the formation of prominence depending upon their morphology, location, activity and the behavior of photospheric magnetic field (e.g., Tandberg-Hanssen 1998; Parenti 2014). Formation of the prominences can be classified according to the location, for example an active region, intermediate, and quiescent prominences. They are associated with the photospheric magnetic field of the Sun. The active region prominences are associated with the strong magnetic field and their polarity inversion lines developing around the sunspots. Active region prominences have large magnetic energy, therefore, they are most unstable one. The quiescent prominences are most stable. They are formed in the higher latitude regions and associated with the weaker photospheric magnetic fields (Priest 1989). The quiescent prominences have the lifetime of a few hours to more than a week with the larger spatial scale. Due to highly dynamic behavior of the quiescent prominences, they show the different irregular motions, structuring, and morphological evolution of the plasma structures, large and small scales flow during their lifetime. The quiescent prominences consist of plasma bubbles which were firstly observed by Stellmacher \& Wiehr (1973) and later they are found to be Rayleigh-Taylor unstable (e.g.; Berger et al. 2010, Ryutova et al. 2010). The quiescent prominences possess some interesting properties like plasma downfall within themselves, plumes, plasma bubbles, plasma blobs, finger structures, and mushroom-like structures within them (e.g.; Berger et al. 2008, 2010; Ryutova et al. 2010; Schmieder et al. 2010). Plumes, plasma bubbles, plasma blobs and other small scale features have never been observed {\bf in} the active region and intermediate prominences. These morphological structures have been used to understand the internal dynamics and magnetic configuration of various prominences, and they may be linked to the different types of the instabilities.\\

The Rayleigh-Taylor instability is initially suggested by Rayleigh (1900) and Taylor (1950) to discuss the interfacial behavior of two fluids when a small perturbation acts on the interface of the fluid. The Rayleigh-Taylor instability occurred when a denser fluid is supported above a low denser fluid and they are accelerated against the gravity. The horizontal component of the magnetic field acting on the interface gives the directionality of magnetic Rayleigh-Taylor instability. HINODE/SOT data has been used to understand the thermal and magnetic causes of the formation of the plume within a hedgerow quiescent prominence (Berger et al. 2008). Later they have observed that these plumes are formed due to the magnetic Rayleigh-Taylor instability (Berger et al. 2010). The three-dimensional numerical simulation has been performed to observe the magnetic Rayleigh-Taylor instability into the filamentary structures (e.g., Jun et al. 1995, Isobe et al. 2005). {\bf Various} magnetic field configurations have been used to study the growth rate of MRT unstable fingers and plasma bubbles (Stones \& Gardiner 2007). The turbulent dark upflow originates from the base of the quiescent prominences, and ejection of the plasma blobs associated with the prominence threads has been observed using SOT data (Hillier et al. 2011a, 2011b). The similar turbulent dark upflows have been observed into the hedgerow quiescent prominence by using SOT data, which were found to be the magnetic Rayleigh-Taylor unstable plumes (e.g., Berger et al. 2008, 2010; De Toma et al. 2008; Ryutova et al. 2010). Later numerical simulation has also been performed to study the nonlinear stability of the Kippehahn \& Schluter model prominence. It is shown that the turbulent dark upflows and the reconnection generated downflows also occur into these prominences. Such plasma dynamics is found to be related with the magnetic Rayleigh-Taylor instability (Hillier et al. 2012a, 2012b).\\

 Van Ballegooijen \& Cranmer (2010) have proposed that the hedgreow prominences consist of large number of thin vertical thraeds and supported by tangled magnetic field. These vertical threads may be formed due to Rayleigh-Taylor instability associated with the tangled field. They consist of the collection of fine thin structures with the width of the {\bf few hundred km to few thousand km} (e.g; Lin et al. 2005; Okamoto et al. 2007.; Chae et al. 2008, 2010). Magnetic field shear affects the growth rate of Rayleigh-Taylor instability but it does not completely supress the instability (Ruderman et al. 2014). Ryutova et al. (2010) have hypothesized that all the plumes in a prominence are not formed due to the magnetic Rayleigh-Taylor instability. Some plumes may also be formed due to the Kelvin-Helmholtz instability by a strong shear flow at the bubble-prominence interface. Multi-mode plume formation is the cause of magnetic Rayleigh-Taylor instability, while single mode plume is formed by the Kelvin-Helmholtz instability. The magnetic Kelvin-Helmholtz unstable vortex-like structures are formed at the surface of fast coronal mass ejection (Foullon et al. 2011, 2013). Ofman \& Thompson (2011) have also observed that the magnetic Kelvin-Helmholtz unstable vortices are developed at the boundary of the erupting region and surrounding corona. The characteristic of prominence-bubble boundary and the effect of the shear flows at the boundary have been observed recently (Berger et al. 2017). The strong shear flows along the bubble-prominence interface leads the coupled Kelvin-Helmholtz \& Rayleigh-Taylor (KH-RT) instability. \\ 

Apart from isolated prominences, the signature of MRT instability has also been observed during their eruption even upto the outer corona and heliosphere. A massive prominence eruption is observed in the solar atmosphere at June 7$^{th}$ 2011, which is associated with an M2.5 solar flares and erupted out from AR 11226 and AR 11227 (Cheng et al. 2012; Yardely et al. 2016). In this prominence, the plasma downfalls have been observed and these falling plasma blobs consist of magnetic Rayleigh-Taylor unstable arcs, fingers, horns and spikes as observed in the lower corona (Innes et al. 2012; Carlyle et al. 2014). These MRT unstable fingers further observed into the intermediate corona in form of MRT unstable finger structures and reach upto the lower interplanetary space where they are fragmented into the plasma spikes (Mishra et al. 2018a,b). Hillier (2018) has reviewed the role of magnetic Rayleigh-Taylor instability into the solar prominences and discussed about the linear phase and nonlinear dynamics of MRT instability. Such instability can be utilized also for the magnetic field estimation of these prominences. \\

\begin{figure*}
\caption{Upper panel: The full FOV of the loop-like prominence observed by SDO/AIA 304 {\AA} on 18$^{th}$ November 2014 after applying multi-Gaussian normalized filter . The sub-region (white-box) consists of an MRT unstable plume. Middle panel: A dark cavity has been evident within the prominence. The spatio-temporal evolution of MRT unstable plume (P1) has been observed. Lower panel: Second plume 'P2' is tracked at different heights and times to estimate the observational growth rate of the MRT instability. Temporal image data in the middle and lower panel show the increment in time from right to left. This is the partial FOV of an eruptive prominence where all MRT instability related plasma proccesses are evolved.} 
\includegraphics[scale=0.8,angle=0,width=18.0cm,height=18.0cm,keepaspectratio]{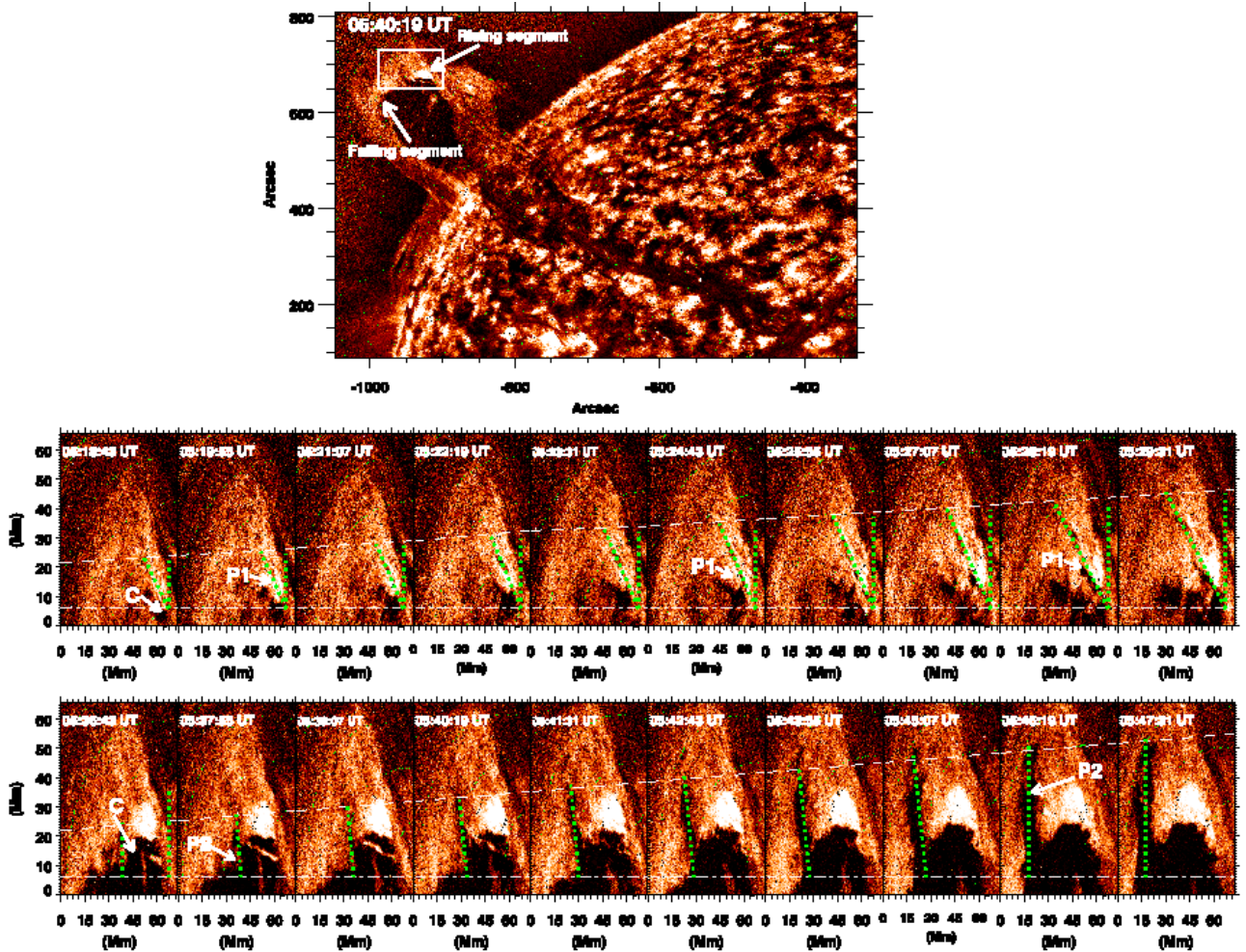}
\end{figure*}
In the present work, to the best of our knowledge, it is first effort to understand the dynamics of the magnetic Rayleigh-Taylor unstable plumes and hybrid KH-RT instability which is associated with the tangled thread in the loop-like eruptive prominence. The details of these MRT unstable multiple localized small-scale plasma structures (two plumes, one thread, shearing motions, multiple plasma blobs) in the eruptive prominence are described in forthcoming sections of the paper. The observations and data analysis have been discussed into the Section 2. Section 3 describes the observational results corresponding to the MRT unstable plumes. The evolution of plasma threads and combined KH-RT instablity is discussed in Section 4. In the final section, discussion and conclusions are outlined. 

\section{Data and Analysis}
The Solar Dynamic Observatory consists of three instruments, the Atmospheric Imaging Assembly (AIA), the Heliospheric Magnetic Imager and Extreme Ultraviolet Variability Experiment (EVE). We analyze the data from the Atmospheric Imaging Assembly (AIA: Lemen et al. 2012) onboard the Solar Dynamic Observatory (SDO). The Atmospheric Imaging Assembly (AIA) have seven extreme ultraviolet (94, 131, 171, 193 , 211, 304, 335 {\AA}), two ultraviolet (1600, 1700 {\AA}) and a visible (4500{\AA}) full disc imager with the 1.5 arcsec spatial resolution and 12 s temporal resolution. We used the 94 {\AA}, 131 {\AA}, 171 {\AA}, 193 {\AA}, 211 {\AA}, 304 {\AA}, and 335 {\AA}  temporal image data of AIA for our analysis. We have selected 3 hour time sequence data starting from 18$^{th}$ November 2014. The basic calibration and normalization of the data were performed by using the Solarsoft IDL routine "aia\_prep.pro". \\

The AIA images are further processed {\bf in order to resolve} the more fine structures, dynamical features on different spatial scales in the eruptive prominence. Each image has convolved with a Gaussian kernel. The convolved images are subtracted from the normal intensity image to obtain the features with better contrast. The same process has been used for five times to measure the uncorrelated noise. The uncorrelated noise is subtracted from the original images. The normalized multiscale Gaussian filter has been used to enhance the fine structures (e.g., Morgan \& Druckm$\ddot{u}$ller 2014; Pant et al. 2015). We select the width of the gaussian filter as [50, 100, 150, 200, 250] pixels. The filtered images are added to the original data. The MRT unstable plumes, prominence thread, vortex formation, and shear motion, all these phenomena {\bf became detectable in the processed images}. Therefore, we perform the analysis by using the multi Gaussian filtered images.  \\  
The thermal properties of the prominence plasma can be understood by estimating the Differential Emission Measure (DEM). We map the DEM with the different temperature from the six AIA filters, i.e, 94 {\AA}, 131 {\AA}, 171 {\AA}, 193 {\AA}, 211 {\AA}, 304 {\AA}. We adopt the method of Cheung et al. (2015) to measure the  emission (EM) from the prominence and surrounding regions. It is based on the concept of the sparse inversion and uses the "simplex" function to minimize the total emission (EM). The sparse inversion technique gives the positive solutions which lying between max(0, I-tol) and (I+tol), where tol is the tolerance into the reconstructed intensities. For the sparse inversion, we divide the range of temperature between log T(K)=4.7-7.2 with 25 temperature bins at $\Delta$log T(K)=0.1 intervals. {\bf Using the DEM analysis, we identify the cool prominence vissible in the temprature range log T(K)=4.7-5.0. Major part of the emission in this temperature range is observed by SDO/AIA 304 {\AA} filter}. The dark plumes and cavity regions are clearly observed in the range of log T(K)=5.8-6.1, which correspond to the AIA 171 {\AA} filter. To estimate the density, we use the estimated total emission coming from different temperature regions :
\begin{equation}
n=\sqrt{\frac{EM}{l}}
\end{equation}
, here n is the number density, EM is the total emission coming from the different temperature bins, and 'l' is the depth of the region from where the emission is occured. The filling factor for the density estimation into the flux rope is assumed to be 1. We assume that the width of the prominence at the prominence-cavity interface is equal to the depth of the prominence (Cheng et al. 2012). The estimated mass density is used for the evaluation of the magnetic field and theoretical growth rate of the MRT instability.\\

 Using above mentioned observational data and various analyses techniques, we have observed a prominence which is in the equilibrium for more than three days. The prominence starts to grow from around 04:30 UT and erupted at 07:38 UT on November 18$^{th}$ 2014. During the rising phase of this prominence, it exhibits the evolution of magnetic Rayleigh-Taylor unstable plumes in its upper part. A cavity has also been developed within the overlying prominence at 05:15 UT, which acts as an initial perturbation. As the perturbation grew, it initiates the development of first MRT unstable plume (P1, Figs.~1-2). Ryutova et al. (2010) has set the criteria for the observed plumes to be Rayleigh-Taylor unstable. These criteria are, multi-mode front, self-similarity and suppression of the regular oscillations of the prominence. We observe the two mode of plumes (P1 \& P2), which show the similarity into the wavelength/width ratio. Strong upflow has {also been} observed into the overlying prominence. These additional perturbations further pass through the prominence-cavity interface and it initiates single mode MRT unstable plume (P2) at 05:36 UT (Fig.~1; bottom panel). The self-similar plume formation (P1 \& P2) shows the linear, and in the later stage the single mode plume formation (P2) exihibiting the non-linear phase of MRT instability as it is converted into the mushroom-like structure in its final stage of the development. The spatio-temporal evolution of MRT unstable plumes is shown in Fig. 1. The MRT instability is destroyed by the shearing and downfall in the prominence-cavity interface. Small-scale vortex-like structures have formed at the interface of downfalling plasma and cavity (Fig.~3). These vortex formation indicates that the MRT unstable plume structures may be converted into the Kelvin-Helmholtz instability and shows the nonlinear phase of an MRT instability. The detailed interpretation is outlined in the subsequent subsection. \\
\begin{figure*}
\begin{center}
\caption{Upper panel: The sub-region of the eruptive prominence using multi-Gaussian normalized filter on AIA 304 {\AA} image. A vertical line is drawn to indicate the initial location of the first plume (P1). The separation between the two consecutive plumes is the characteristic wavelength of an MRT instability. Lower panel: The Differential Emission Measure (DEM) of the same sub-region to observe the thermal structure of this MRT unstable prominence.}
\includegraphics[scale=0.8,angle=90,width=12.0cm,height=8.0cm,keepaspectratio]{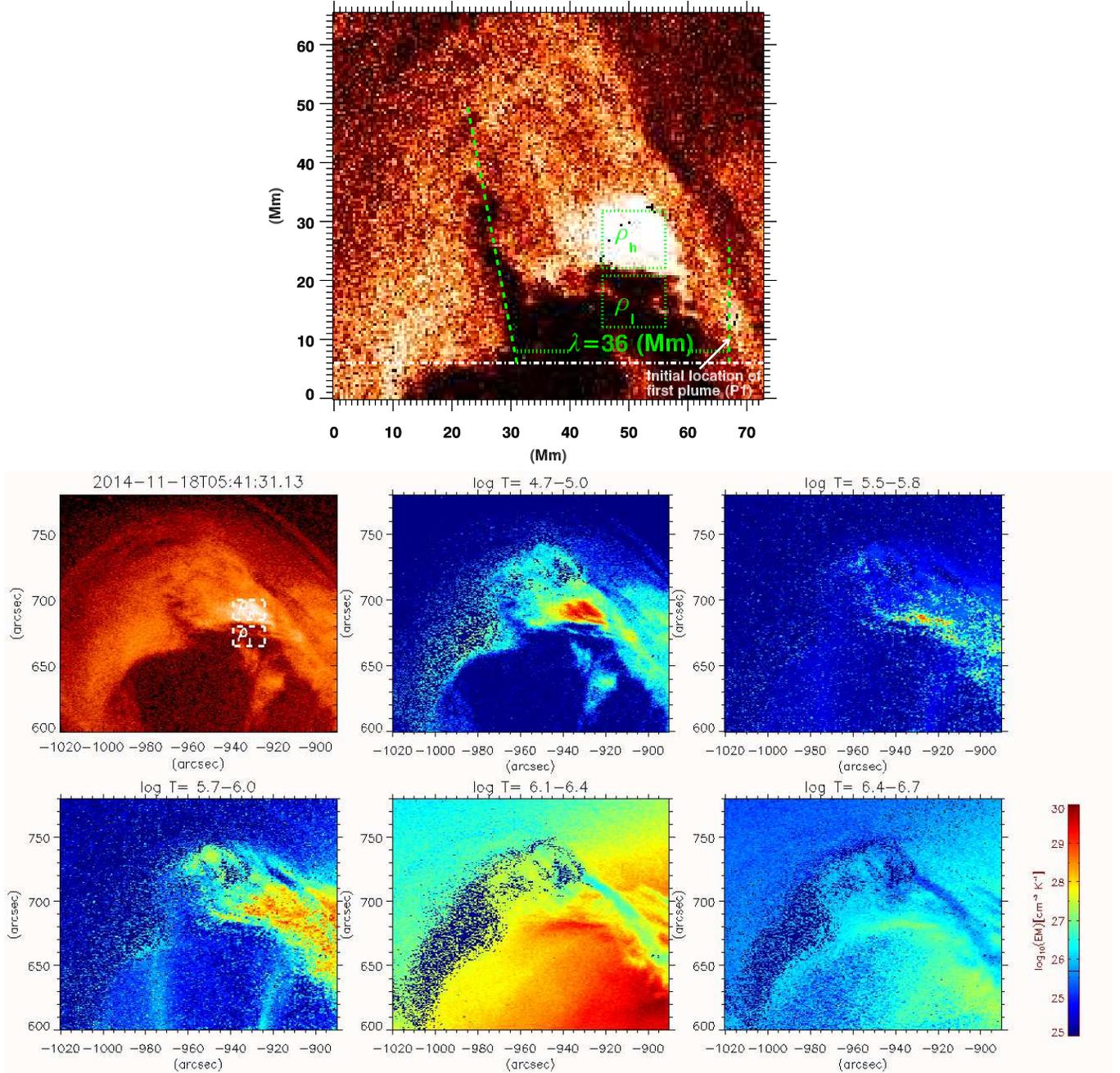}
\vspace {-0.9cm}
\end{center}
\end{figure*}
\begin{figure*}
\begin{center}
\includegraphics[scale=0.8,angle=0,width=18.0cm,height=11.0cm,keepaspectratio]{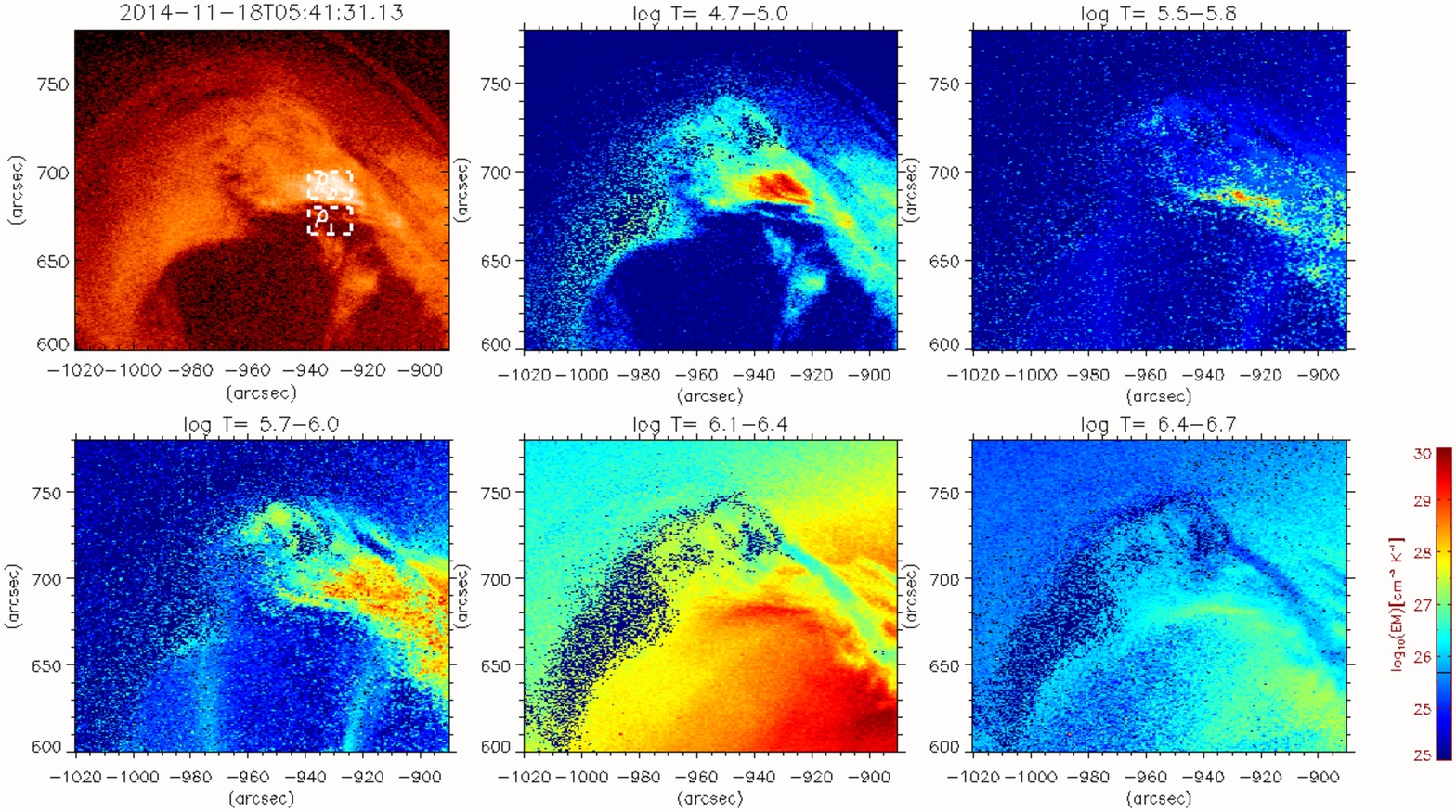}
\end{center}
\end{figure*}
We observed that in the rising plasma segment, a tangled plasma thread passes through the overlying prominence at 06:43 UT (Fig.~4). The tangled vertical thread is associated with Rayleigh-Taylor instability (van Ballegooijen \& Cranmer 2010). The tangled thread passes into the overlying prominence with velocity $\approx$220 km s$^{-1}$. In the tangled thread, the bright location indicates a dip (Fig.~4). The thickness of the thread has been observed as $\approx$3.5 Mm. As it passes through the overlying prominence, a cavity has been developed along the length of the thread and later it is responsible for the shearing into the thread like structures (Figs.~5-7). Due to shearing motion at the interface, small-scale vortex-like structures move along the thread with the lifetime of 150-200 s. These vortex like structures may be formed due to the Kelvin-Helmholtz instability at the interface of the prominence and cavity. In the later time, more than five plasma blobs have been observed at the same interface (Fig.~8). These silent features of the observed highly localized plasma processes (i.e, plumes, thread, blobs) in different evolutionary stages of an eruptive prominence are explained with their theoretical interpretation in Sect. 3-4.\\

\section{Observational Results}
\subsection{ Formation of MRT Unstable Plumes}
As abbreviated above, we observed a loop-like prominence eruption on 18$^{th}$ November 2014 during 04:00 UT to 07:50 UT by SDO. Before going to the eruptive stage, this quiescent prominence remains in the stable state for more than three days. We used the high-resolution and high-cadence multi-wavelength data (94 {\AA}, 131 {\AA}, 171 {\AA}, 193 {\AA}, 211 {\AA}, 304 {\AA}, 335 {\AA}) of Atmospheric Imaging Assembly (AIA) onboard SDO (Lemen et al. 2012) to observe the dynamics of the magnetic Rayleigh-Taylor unstable plumes and plasma threads. Prominence plumes are dark structures in the cool temperature filters of AIA and they appeared bright in form of hot structures in compratively high temperature filters. The plume-like structure is generated from the large-scale cavity or bubbles which are evolved within the prominence. They pass through the overlying prominence before they get fragmented. The plumes and plasma bubbles were first observed by Stellmacher \& Wiehr (1973) and later it was hypothesized that these plumes are generated due to the magnetic Rayleigh-Taylor instability (Ryutova et al. 2010). We observed a loop-like prominence which consists of the MRT unstable plume like developments into its upper part (Fig.~1). We observed that the small-scale bubbles started to grow on 18$^{th}$ November 2014 after 05:00 UT. The source of these bubbles are not clear. Initially, these bubbles are developed in the horizontal as well as vertical directions both. Small scale perturbations are developed at the boundary of the cavity and overlying prominence. We observed that some of these perturbations further developed on a small-spatial scale. One of the perturbation becomes unstable at 05:15 UT and it passes through the overlying prominence with dark upflows (Fig.~1). The magnetic Rayleigh-Taylor instability is appeared when a contact of discontinuity is formed along the magnetic interface where a denser fluid is lying over the less denser fluid and accelerate against the gravity. We observed that the cavity and the dark upflows generated within the overlying prominence are lying in the temperature range of (6.3-12.5)$\times 10^{5}$ Kelvin which is 12-25 times hotter than the surrounding prominence plasma (Fig.~2). The hotter plasma is lying below the cooler plasma and hence it may act as the source of buoyancy. Due to the buoyancy force, the dark plume like structures are lifted up with the vertical uplflows. The interface of the prominence-cavity system supports the magneto-thermal convection into the solar atmosphere (Berger et al. 2011). The bright and denser region is associated with the overlying prominence and the dark and less denser cavity is associated with the ambient coronal plasma (Fig.~2). The denser plume and prominence region is lying at different temperatures, therefore, they are magnetically insulated because thermal conduction in the corona occurs along the field lines (Innes et al. 2012; Priest 2014). A magnetic interface is presented between the hot plume and cool prominence region. The density gradient is working towards the Sun center along the direction of gravity. The magnetic tension  component of the Lorentz force may accelerate the dark plumes against the gravity and density gradient. Therefore, the dark plumes have satisfied the criteria of magnetic Rayleigh-Taylor instability. The lifetime of the dark plume (P1) is $\approx 14$ minutes and it propagates with the constant velocity of $\approx$ 35 km s$^{-1}$ in the overlying region before it is fragmented (Fig.~1, middle panel). The width of the plume (P1) lies between 3-7 Mm (Fig.~1). The main characteristic of the magnetic Rayleigh-Taylor instability is that the formation of multiple self-similar plumes occur in its linear phase (Ryutova et al. 2010).\\
The cavity further grows and it gains the semi-spherical shape (Fig.~1; bottom panel). Additional perturbations have been developed at the cavity-prominence interface, which further trigger the second plume (P2) (Fig.~1; bottom panel). We observed that this plume also satisfies the MRT instability criteria. The development of the second plume (P2) indicates the transformation from the linear phase to the nonlinear phase of MRT instability. We observe that the plumes has shown the vertical upflows in the linear regime of MRT instability during 05:36 UT to 05:43 UT. At the boundary of the second plume (P2), regular dark and bright oscillatory patterns are presented. These regular oscillatory patterns indicate that the multiple self-similar RT unstable plumes may be evolved. After 05:43 UT, we observed that the interface of the dark plume (P2) and prominence became smooth and it may show the quasilinear phase of MRT instability. Later after 05:46 UT, it transforms into a single mode plume and exhibits the nonlinear phase of MRT instability. We observe that this plume takes the shape of mushroom cap structure in its final stage of the development before going to the stable mode (Fig.~1; bottom panel). The nonlinear explosive phase of the Kelvin-Helmholtz instability and the mushroom-like structures via nonlinear MRT instability are characterized by the single mode plume formation (Ryutova et al. 2010). Therefore the final stage of the plume (P2) formation exhibits the nonlinear phase of the instability where the single mode plume is dominating by making mushroom cap-like structure (Fig.~1, lower panel). The lifetime, width and the upflow velocity of the single mode plume is greater then the multi-mode plume formation. We have observed that the lifetime of plume (P2) is greater than 40 minutes, the maximum width is 14-16 Mm. It grows with the constant upflow velocity of 49 km $s^{-1}$ (Fig.~1).

\subsection{Estimation of Theoretical and Observational Growth Rate of MRT Unstable Plumes}
We identify two MRT unstable plumes (P1 \& P2) which propagate in the form of dark upflows into the overlying prominence with the velocity of $\sim$35 and $\sim$49 km s$^{-1}$ respectively. We observe the spatial and temporal evolution of the MRT unstable plumes (Fig.~1). The growth rate of the MRT instability during its linear and quasilinear phase at height (h) and time (t) (Ryutova et al. 2010) is given by,
\begin{equation}
\gamma_{Obs}=\frac{1}{(t_{2}-t_{1})}ln(\frac{h_{2}}{h_{1}})
\end{equation}
, where $h_{1}$ and $h_{2}$ are the height of the plumes at time $t_{1}$ and $t_{2}$. A well evolved plume has been first appeared on 18$^{th}$ November 2014 at 05:16 UT. We track the growth of the plume at a different height with time (Fig.~1). The temporal and spatial evolution of the MRT unstable plumes have been observed for the estimation of the observational growth rate. The estimated observational growth rate of the two MRT unstable plumes are listed in Table 1. The average value of the observatioanl growth rate for plume (P1) is 1.32$\pm$0.29$\times$ 10$^{-3}$ s$^{-1}$ and for plume (P2) is 1.48$\pm$0.29$\times$ 10$^{-3}$ s$^{-1}$. We observe that the observational growth rate for both the MRT unstable plumes decrease with the increament in their heights. \\

We compare the observational growth rate of the MRT unstable plumes with the theoretical growth rate of the MRT instability. If $\gamma_{th}$ is the theoretical growth rate of the MRT instability, $\rho_{h}$ and $\rho_{l}$ indicate the mass density within the bright dense region (plumes here) and dark below lying cavity region respectively at the prominence-cavity interface, the linear stability theory (Chandrasekhar 1961; Ryutova et al. 2010; Priest 2014) describes the growth rate of the MRT instability as, 
\begin{equation}
\gamma_{th}=\sqrt{\frac{2\pi g}{\lambda}A[1-{\frac{B^{2}cos^{2}\theta}{(\rho_{h}-\rho_{l})g\lambda}}]}
\end{equation}
\begin{figure*}
\caption{ Top panel: The spatio-temporal varition of the full FOV of the overlying prominence which collapses by plasma downfall. Two white boxes 'a' and 'b' indicate the region where KH unstable vortices and rolled plasma structures are formed. The evolution of these vortices and rolled plasma structures are shown in the lower panels. An animation has also been added to show the plasma dynamics during the downfall.}
\includegraphics[scale=0.8,angle=0,width=18.0cm,height=18.0cm,keepaspectratio]{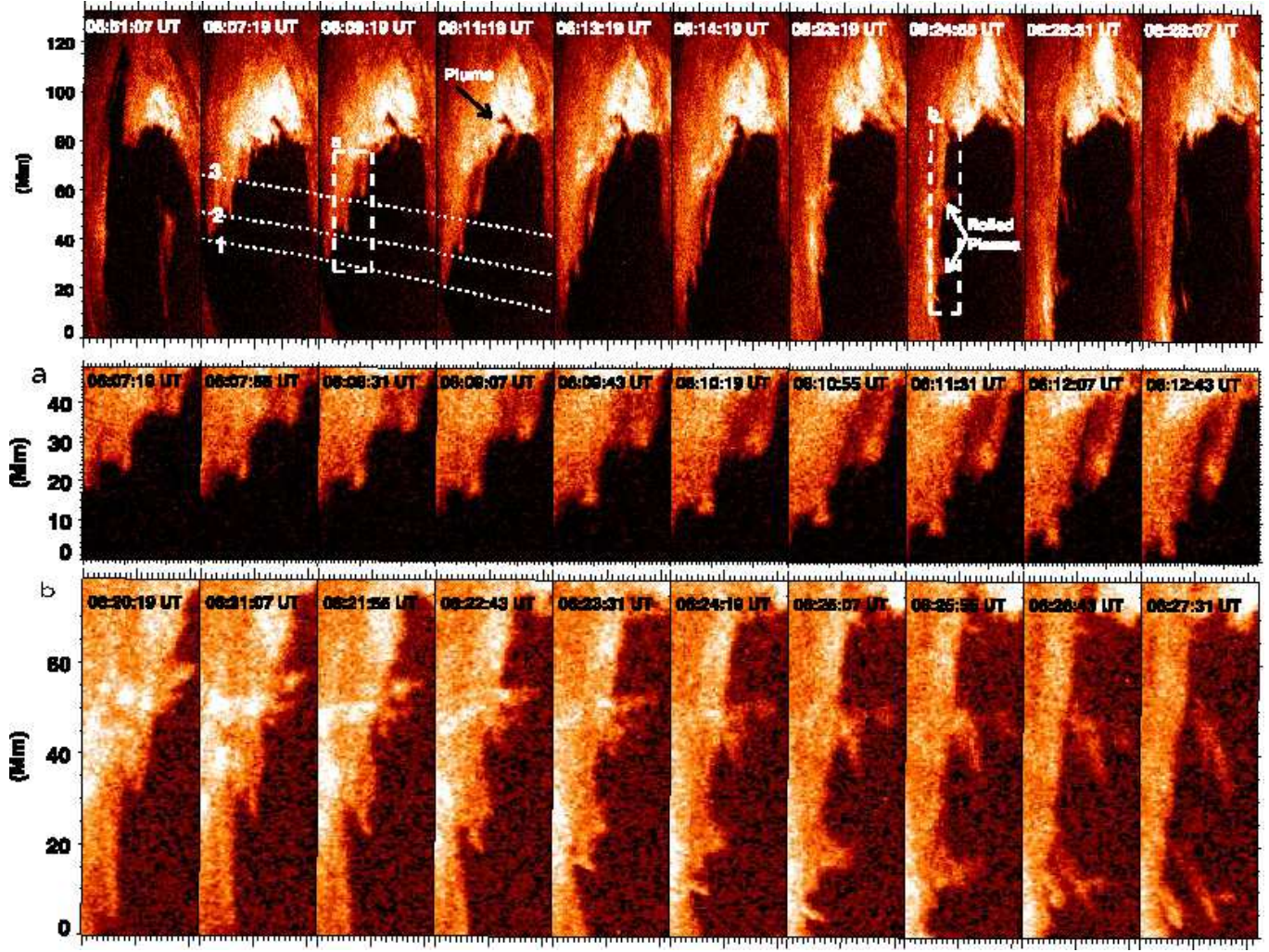}
\end{figure*}
here g is the acceleration due to the gravity of the Sun, B is the horizontal component of the magnetic field, $\lambda$ is the  characteristic wavelength of the MRT instability, A is Atwood number which is given by A=$\frac{\rho_{h}-\rho_{l}}{\rho_{h}+\rho_{l}}$, and $\theta$ is the angle between the magnetic field (B) and wave vector of the MRT unstable perturbations. The magnetic field component of the growth suppresses the instability. Therefore, we can say that as the plumes pass through the higher height, the magnetic field component reduces the growth of MRT instability. We observe that the theoretical growth rate contains all the measurable parameters. \\
 
\begin{table}
\caption{Projected height estimation of the plumes (P1 \& P2) w.r.t. normalized height. The normalized height is considered with respect to the position of cavity 'C' situated at 471 Mm from Sun's center. The observational growth rate of MRT instability is estimated by tracking the height of the observed plumes at different time. }
\label{table:1}
\begin{center}
\begin{tabular}  {|p{3.4cm}|p{3.4cm}|p{3.4cm}|p{3.4cm}|p{3.4cm}|}
\hline
Projected Height (Mm) of the Plume (P1) with Respect to Postion of the Cavity "C" & Projected Height (Mm) of the Plume (P2) with Respect to Postion of the Cavity "C" & Time interval (t$_{2}$-t$_{1}$) & Observed growth rate for the plume (P1) $\gamma_{Obs1}=\frac{1}{(t_{2}-t_{1})}ln(\frac{h_{2}}{h_{1}})\times 10^{-3} s^{-1} $ & Observed growth rate for the plume (P2) $\gamma_{Obs2}=\frac{1}{(t_{2}-t_{1})}ln(\frac{h_{2}}{h_{1}})\times 10^{-3} s^{-1}$  \\
\hline
16.5 & 18.3 &0 &0 & 0 \\
\hline
19.0 & 21.6 & 72 &1.96$\pm$0.29 & 2.30$\pm$0.29 \\
\hline
21.5 & 24.9 & 72 & 1.72$\pm$0.21 & 1.97$\pm$0.20 \\
\hline
24.0 & 28.1 & 72 & 1.53$\pm$0.16 & 1.68$\pm$0.15  \\
\hline
26.5 & 31.4 & 72 & 1.38$\pm$0.13 & 1.54$\pm$0.11  \\
\hline
29.0 & 34.7 & 72 & 1.25$\pm$0.10 & 1.39$\pm$0.09  \\
\hline
31.5 & 38.0 & 72 & 1.15$\pm$0.08 & 1.26$\pm$0.07  \\
\hline 
34.0 & 41.3 & 72 & 1.06$\pm$0.07 & 1.16$\pm$0.06  \\
\hline
36.5 & 44.6 & 72 & 0.98$\pm$0.06 & 1.06$\pm$0.05  \\
\hline
39.0 & 47.9 & 72 & 0.92$\pm$0.05 & 0.99$\pm$0.04  \\
\hline
\end{tabular}
\end{center}
\end{table}
 We adopt the sparse inversion code to estimate the mass densities within the plume regions and below lying dark coronal region (Cheung et al. 2015). The width of the prominece at the prominence-cavity interface is $\sim$75 Mm. The total Emission Measure (EM) coming from the bright region and dark cavity region lying below of it have found to be 3.2$\times$10$^{29}$ cm$^{-5}$ and 1.4$\times$10$^{29}$ cm$^{-5}$ respectively along the prominence body. The plasma density (n=$\sqrt{\frac{EM}{l}}$) associated with these two regions have been estimated as 6.5$\times$10$^{9}$ cm$^{-3}$ and 4.3$\times$10$^{9}$ cm$^{-3}$ respectively. The assumed composition of the prominence is 45\% H$^{\circ}$, 45\% H$^{+}$, 9\% He$^{\circ}$, 1\% He$^{+}$ (Labrosse et al. 2010; Gilbert et al. 2011). The estimated mass densities for the bright dense region $\rho_{h}$ and the dark cavity region $\rho_{l}$ have found to be 1.4$\times$ 10$^{-14}$ and  9.3$\times$ 10$^{-15}$ gm cm$^{-3}$ respectively (Fig.~2).  An updated version of the sparse inversion code is "sparse code new", which is useful to measure the total emission of the hotter plasma (Su et al. 2018). This new updated version introduces the sparse inversion code and mapped the thermal plasma from $\sim$0.3 to $\sim$30 MK. It is important particularly for flare initiation, current sheets, cusps, super-arcade downflows, shocks, active regions, EIT waves, etc. In the present case, we observe that most of the emissions are coming from the lower temperature (304 \AA) for the prominence. The emission from 304 $\AA$ filter is opticaly thick and with high opacity as compared to the other six filters of SDO/AIA. The higher opacity may be the cause of the saturation of the continuum absorptions and therefore, slight underestimation of the mass density may be observed (Heinzel et al. 2008; Gilbert et al. 2011). In the given observational base-line, therefore, we utilize the method of Cheung et al. (2015), which gives almost consistent results as derived from the Su et al. (2018).\\

From Equation 3, when $\theta$=0, i.e. for the condition where the magnetic field is oriented in the same direction for the dense region and below lying less denser region, the maximum growth rate could be attained (Chandrasekhar 1961; Ryutova et al. 2010; Priest 2014). Therefore, for the maximum growth rate, $\gamma_{th}$=0,
\begin{equation}
-gk{\frac{(\rho_{h}-\rho_{l})}{(\rho_{h}+\rho_{l})}}+{\frac{B^{2}k^{2}cos^{2}\theta}{2\pi(\rho_{h}+\rho_{l})}}=0
\end{equation}
, where k is the wave vector associated with MRT instability. The instability is occurred when k$<$k$_{c}$, which is related with the critical wavelength $\lambda_{c}$. The critical magnetic field (B$_{c}$) is given by,
\begin{equation}
B^{2}=g\lambda_{c} (\rho_{h}-\rho_{l})
\end{equation}
We observe that the bright denser regions are heavier than the less denser dark regions, so we take $\frac{\rho_{h}}{\rho_{l}}>$1. The Alfven velocity for the rising MRT unstable plume is (Innes et al. 2012),
\begin{equation}
B^{2}=4\pi\rho_{h}V_{A}^{2}
\end{equation}
Therfore we obtain that,
\begin{equation}
V_{A}=\sqrt{\frac{\lambda_{c}g}{4\pi}}
\end{equation}
\begin{figure*}
{\caption: Top panel: Full FOV of multi-Gaussian normalized and fitted SDO/AIA 304 {\AA} filter, which consists of Rayleigh-Taylor unstable prominence thread. Middle panel: The sequence of images of the sub-region in SDO/AIA 304 filter showing the temporal evolution of the plasma thread. Lower panle: The V-shaped dip as evident in the lower panel in the SDO/AIA 171 {\AA} filter.  }
\includegraphics[scale=0.8,angle=0,width=18.0cm,height=18.0cm,keepaspectratio]{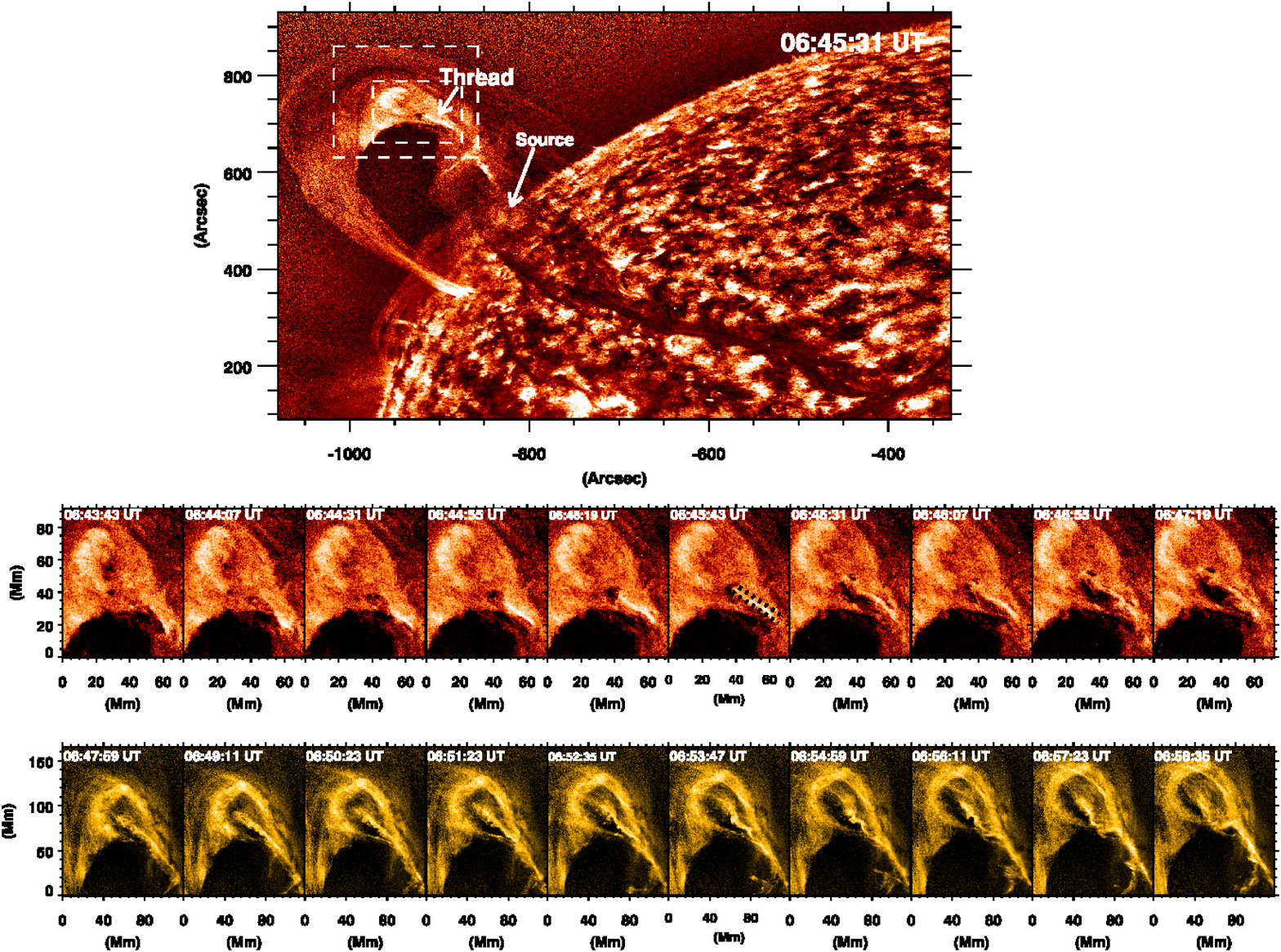}
\end{figure*}

The separation between two consecutive MRT unstable plumes (Fig.~2) gives the characteristic wavelength $\lambda$=36 Mm. Analogous to this wavelength, the Alfven velocity (V$_{A}$) is $\approx$ 28 km s$^{-1}$. We have estimated the magnetic field (B)$\approx$1.32 G. Therefore, to estimate the theoretical growth rate, we take, $\rho_{h}$= 1.43 $\times$10$^{-14}$ gm cm$^{-3}$, $\rho_{l}$= 9.32$\times$ 10$^{-15}$ gm cm$^{-3}$, Atwood number. (A)=0.21, Solar gravity (g)=274 m s$^{-2}$, and characteristics wavelength ($\lambda$)= 36 Mm. We observed that the first plume propagates at an angle $\theta_{1}$=48$^\circ$. For the second plume (P2), $\theta_{2}$ is 82$^\circ$ from the top of the cavity, where the plume originates in the projected plane. The theoretical growth rate of the MRT instability is estimated as $\gamma_{th}$=1.95$\times$ 10$^{-3}$ s$^{-1}$, which is in good agreement with the observed growth rate (Table 1). \\
We observe that after the full development of the second MRT unstable plume (P2), the downfall starts into the falling segment. The full growth of the plume (P2) is occurred at 05:50 UT, where it gains the mushroom cap-like structure (Fig.~3). The MRT unstable plumes try to attain stability. Series of the time-sequence images have been used to study the spatial variation of the downfalling plasma (Fig.~3, top panel). The additional perturbations are continuously developed at the interface of the prominence and cavity. To gain the stability, the loop-like prominence releases the shear at the prominence cavity interface (Fig.~3). After 06:03 UT, some of the perturbations are developed non-linearly and converted into the MRT unstable plume and some rolled-up structures at the prominence-cavity boundary (Fig.~3; boxes 'a' \& 'b'). At the boundary of the falling plasma and cavity, three small scale rolled-up plasma substuctures have been observed (see also the animation1.mpg). These rolled substructures are formed at the regular intervals. The velocity of the leading edge of the downfalling plasma and two rolled-up structures which are indicated by 1, 2 and 3 have found to be 70, 58 and 55 km s$^{-1}$ respectively. It should be noted that under the given spatial resolution and emissions of AIA 304 $\AA$, the fine structures related to the rolling in these vortices are observed. It is clear from the Fig.~3 and related animation that the central bright core within these small-scale structures are rolling and compratively less intense plasma around the core is trying to envelope and bend (panels 'a' \& 'b'; animation1.mpg). However, we wish to point out that this is more a qualitative description based on the obtained 304 $\AA$ images as observed by AIA. We need more high resolution observations to resolve the fine structured vortex motions at prominence-cavity interface. Later, few rolled plasma structures are observed at the prominence-cavity interface (Fig.~3; lower panel). These plasma structures may be formed due to the Kelvin-Helmholtz instability. Therefore, we conjecture that the MRT unstable plumes attain the stability by Kelvin-Helmholtz vortex formation. We have added the animation in the support of the observed rolled-up structures and related plasma dynamics as seen in Fig.~3. 
\begin{figure*}
{\caption: The sequence of the images of SDO/AIA 171 {\AA} shows the boundary of the prominence-cavity, which became unstable. The middle panel shows that two bigger plasma bubbles are observed. Later they may be responsible for the shear motion at the interface. The shear motion initiates the Kelvin-Helmholtz unstable vortex formation and Rayleigh-Taylor unstable bubbles are also formed at the boundary.} 
\includegraphics[scale=0.8,angle=0,width=18.0cm,height=18.0cm,keepaspectratio]{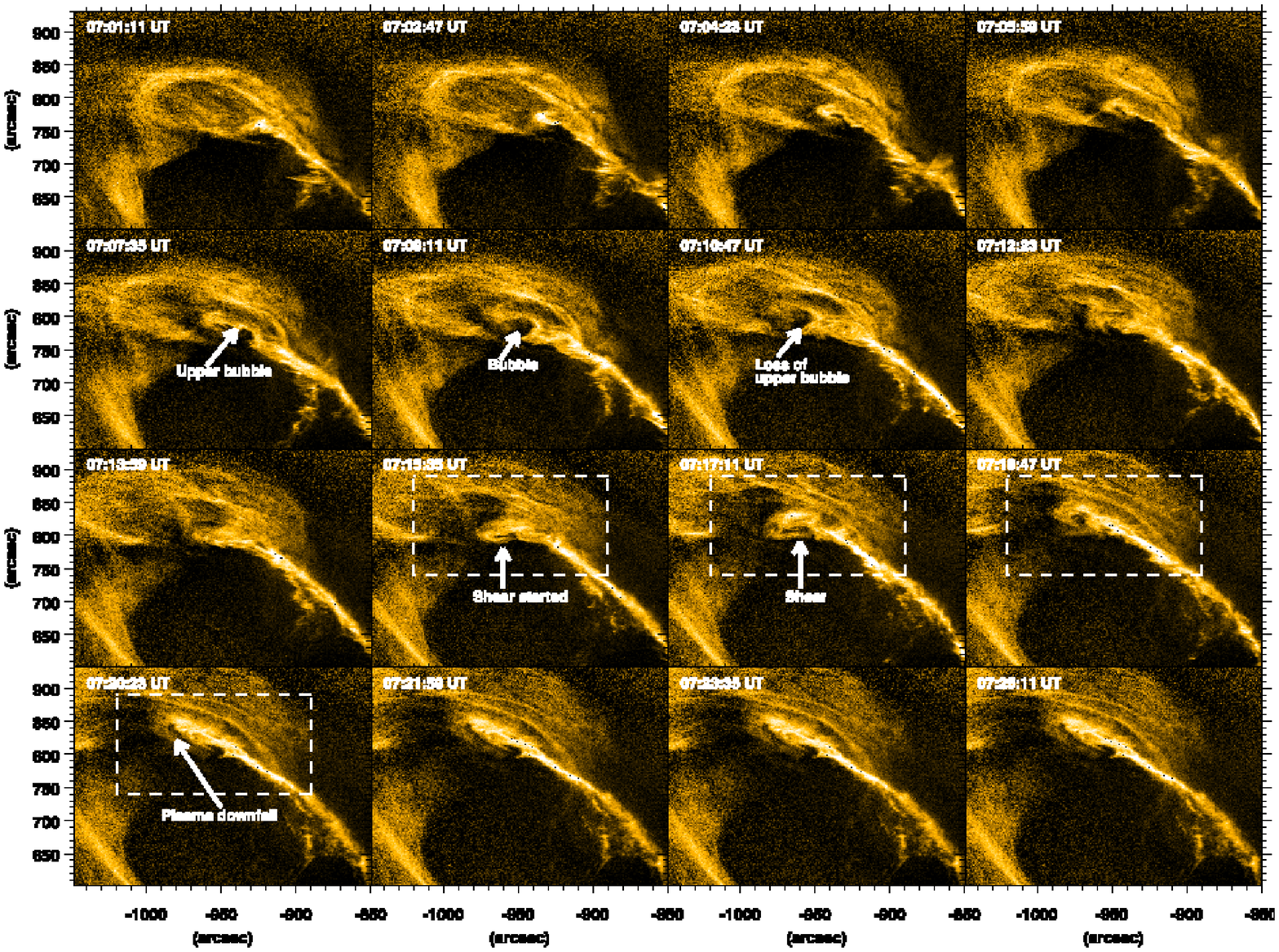}
\end{figure*}

\section{Formation of Tangled Threads and Plasma Bubbles due to the hybrid KH-MRT instability}
\subsection{Tangled Threads}
\begin{figure*}
{\caption: The region of interest (ROI; white box in Fig.~5) is displayed in the SDO/AIA (171+304){\AA} composite images. The composite images observed the characteristic of the bubble-prominence boundary. Shear motion, plasma downfall, MRT unstable bubbles associated with the Rayleigh-Taylor unstable thread are indicated by arrows at the interface. The animation2.mpg shows the complete evolution of shearing, bubbles formation, MRT unstable bubbles and collapse of the bubbles at the prominence-cavity interface.}
\includegraphics[scale=0.8,angle=0,width=18.0cm,height=18.0cm,keepaspectratio]{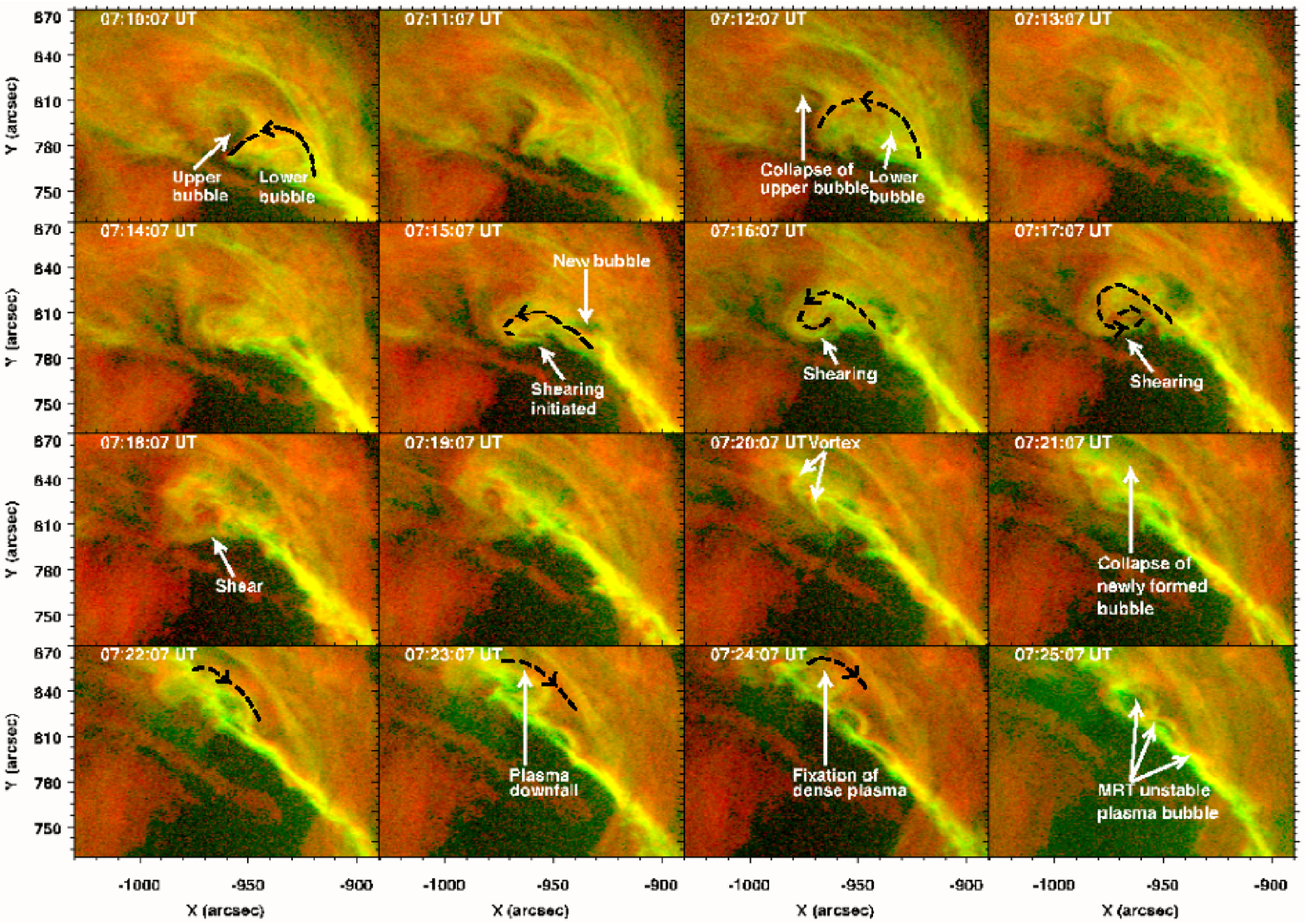}
\end{figure*}
We observed that after one hour of the evolution of MRT unstable plumes, a thin bright thread is observed in the rising segment of the loop-like prominence. The thread first appeared on 18$^{th}$ November 2014 at 06:44 UT (Fig.~4). The main source of the thread is not clear on the solar disk, because the lower part of the prominence is partially covered by the eruption site. The typical lifetime for the thread is 8-10 minutes with the maximum thickness of 3.5 Mm (Fig.~4, middle panel). A dark cavity region has been observed between the top of the thread and overlying prominence. The thread drags the cavity region, and it is trapped within the overlying prominence (Fig.~4, middle panel). The random orientation of field lines and V-shaped structures are observed at the prominence-cavity interface (Fig.~4, bottom panel). The prominences are supported by the pressure of the sheared magnetic field. It acts as the tenuous fluid region and shows the buoyant nature. If a tenuous fluid supported the denser fluid against the gravity it becames Rayleigh-Taylor unstable (Chandrasekhar 1961). The observed thread satisfies the similar constraints and may be subjected to the Rayleigh-Taylor instability (Van Ballegooijen \& Cranmer 2010). The cool plasma is collected into the dip regions of the tangled field. The temporal and spatial evolution of the tangled plasma thread has observed (Fig.~4).
 \begin{figure*}
{\caption: The region of interest (ROI; white box in Fig.~5) is displayed using SDO/AIA 304 {\AA} images. MRT unstable bubbles, shear motion, and KH unstable vortex have been indicated by arrowa at different time-epochs of the prominence-cavity boundary. The animation3.mpg shows the complete behaviour of the prominence-cavity boundary and different morphological structures.}
\includegraphics[scale=0.8,angle=0,width=18.0cm,height=18.0cm,keepaspectratio]{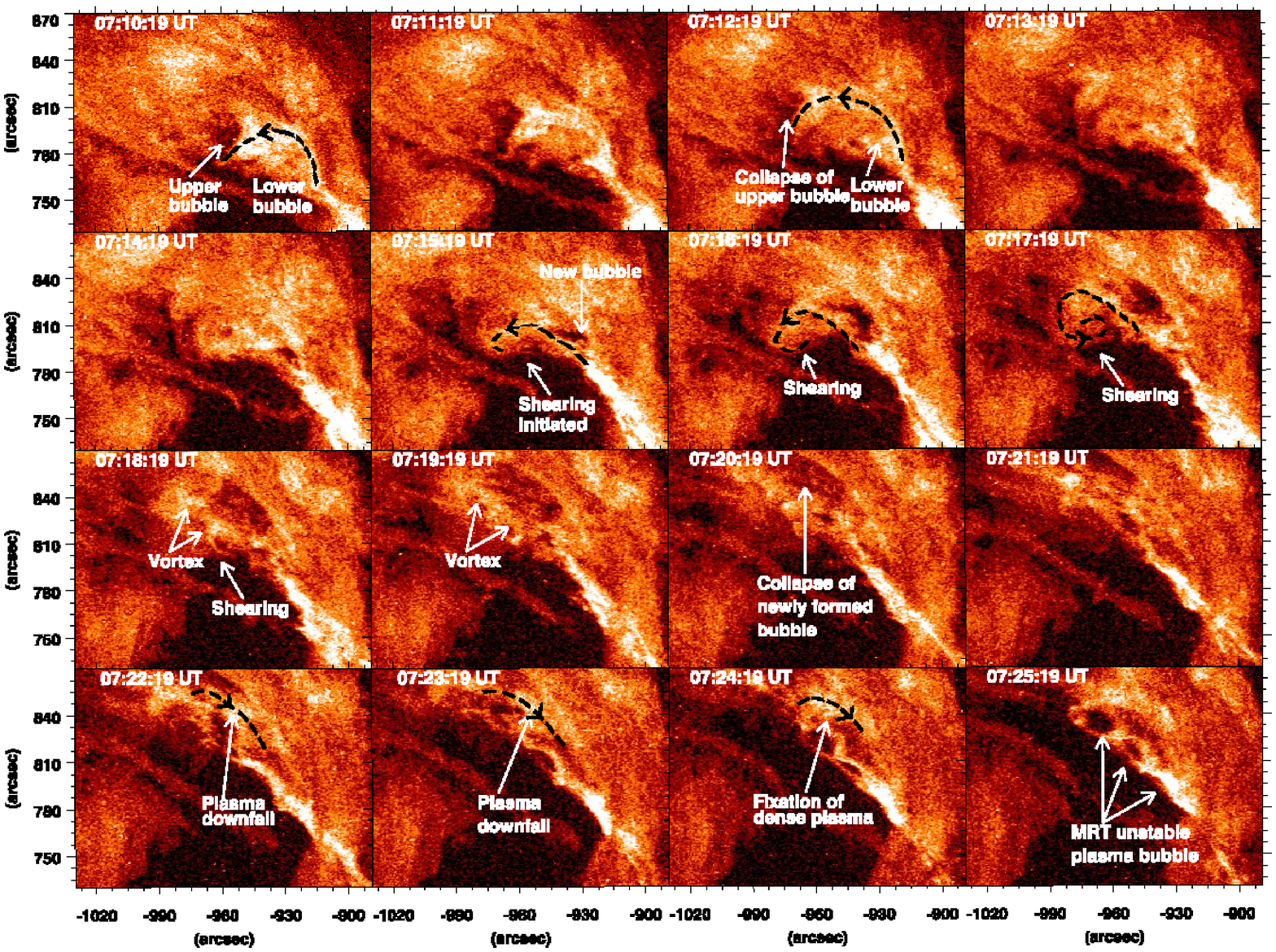}
\end{figure*}   
\subsection{Evolution of the Hybrid KH-RT Instability}
In this section, we discuss the possibility of the hybrid Kelvin-Helmholtz \& Rayleigh-Taylor (KH-RT) instability at the prominence-cavity interface. We observe that as the tangled plasma thread passes through the overlying prominence, small-scale bubbles have been developed between the thread and surrounding prominence. The loop-like prominence eruption is triggered by an instability which is associated with the shear on the prominence-cavity interface (Fig.~6, 7, animation2.mpg, animation3.mpg). The shear developed at the interface is diffused and rapid, therefore the analysis of the shear and instability is very difficult. The optical flow measurements like, Local Correlation Tracking (LCT), Differential Affine Velocity Estimator (DAVE) and Non-linear Affine Velocity Estimator (NAVE) may not be useful in the present observation to understand the fine structures as the shearing developed far off the limb (Chae \& Sakurai 2008). In this section, we do not perform any quantitative analysis, but we observe the shear qualitatively by analyzing the time-series of the imaging observations.\\
 
The thread-like structures appear on 18$^{th}$ November 2014 at 06:43 UT and interact with a leg (northward lying) of the overlying prominence (Fig.~4). Two eruption sites have been observed which continuously erupted the plasma along the thread and these plasma interact with the one leg of the overlying prominence. The eruption sites are not clearly observed on the solar disk as they are partially covered by the prominence leg. The observed plasma thread is trapped into the upper loop-like prominence and bigger cavity region. The tangled thread has dip region which is observed as V-shaped structures containing more prominence plasma. The loop-like prominence and thread consist of the tangled magnetic field. Due to the relative motion of these two, the field lines stretch into the vertical direction. Therefore, the shear increases with time at the prominence-cavity interface (Figs.~5, 6, 7, animation2.mpg \& animation3.mpg). The spatial dynamics of the cavity-prominence boundary has been observed by using the sequence of images (Figs.~5, 6, 7, animation2.mpg \& animation3.mpg). We observe that a new cavity/bubble has been developed at the prominence-cavity interface after 07:07 UT. This cavity has been dragged by the dense plasma material, which is linked with the two eruption sites and trapped between the thread and overlying prominence. The dense region coming from the eruption site fills the cavity and it may change the structure of cavity when it interacts with the overlying thread and prominence. The observed cavity region is associated with the low dense and hotter plasma materials and excess of emission from around 1 MK coronal temperature (Berger et al. 2011; Hillier 2018). They are surrounded by the cooler and dense prominence plasma (Fig.~6; animation2.mpg, animation3.mpg). The hotter and low dense plasma is lying below the cooler and higher dense prominence plasma and hence it may act as the source of buoyancy and density inversion. Therefore, the observed plasma bubbles are magnetic Rayleigh-Taylor unstable at the prominence-cavity interface. After 07:10 UT, we observe the complex configuration of the prominence-cavity interface and collapse of the cavity at the same interface. The collapse of the cavity may trigger the shear flows at the interface. To observed the shear flow qualitatively, we have selected the Region of Interest (dotted white box in the bottom panel; Fig.~5). The shear flow is very diffused and rapid, therefore we use the array of images having duration 15 minutes with 60 s cadence. The shear flow collapses after few minutes by plasma downflow near the prominence-cavity boundary. The plasma downflow and shear flows may trigger the instability at the cavity-prominence boundary. The shear flow fixes the higher dense plasma on the top on the thread and small-scale plasma bubbles are formed at the boundary of the prominence-cavity. We track these plasma bubbles on different heights and time by taking the slits at a different particular heights (Fig.~8, top panel). We observed that MRT unstable plasma bubbles grew with time and got a regular oscillatory pattern with the approximate period of 124$\pm$18 s (Fig.~8; middle and lower panel). An oscillatory pattern at the interface may indicate the bumps of the plasma at the interface. These bumps may be developed due to the Kelvin-Helmholtz instability, caused by shear flow at the prominence-cavity interface. The observed plasma bubbles are magnetic Rayleigh-Taylor unstable and Kelvin-Helmholtz unstable simultaniously at the interface. Therefore, we discussed the possibility of hybrid KH-RT instability at the prominence-cavity interface. Some theoretical works were already done on the coupled KH-RT instability (c.f., Zhang et al. 2005; Olson et al. 2011; Ye et al. 2011). Berger et al. (2017) have used the HINODE/SOT data to observe the dynamics of the prominence-cavity boundary. He reported that the quiescent prominence-cavity interface dynamics shows the evolution of the coupled KH-RT instability which was the first observational proof of the hybrid KH-RT instability.\\
 The different plasma bubbles move with different speed on the interface. In the upper part of the eruption, shearing start due to the relative motion of the erupted plasma and overlying prominence. Due to their relative motions and different velocity of plasma bubbles, they are trying to merge within itself (Fig.~8). We have estimated the theoretical merging time for these plasma bubbles which is given by (Ryutova et al. 2010),
\begin{equation}
\tau_{th}=\frac{\Delta U}{0.1 Ag}
\end{equation}
, where '$\Delta U$' is a velocity difference between two consecutive bubbles, 'A' is Atwood number and 'g' is the acceleration due to gravity.  The density within the bright plasma blobs is ($\rho_{h})$=1.65$\times$ 10$^{-14}$ gm cm$^{-3}$. In the below lying dark region ($\rho_{l})$=0.92$\times$ 10$^{-14}$ gm cm$^{-3}$. These densities have been estimated by measuring the total emissions (Cheung et al. 2015). Using the measured velocity difference $\Delta U$=12 km s$^{-1}$, Atwood number (A)= 0.29, we get the theoretical merging time ($\tau_{th}$)= 24 min 30 s. The observational merging time is approximately 17 min 40 s, which can be observed from Fig.~8. The theoretical merging time is reasonably comparable to the observational merging time of the plasma bubbles. \\
The perturbations coming from the two eruption sites and the associated shear flow may be the main cause of hybrid KH-RT instability. During the growth of this instability, the overlying prominence also becames unstable. As the merging of the plasma blobs is completed, the overlying prominence became unstable and erupted towards the outer corona. Therefore, we could assume that the hybrid KH-RT instability may be the main cause of the eruption of the overlying prominence and generation of the associated CME. The lower segment of the prominence thread consists of the bulky plasma material, which is further observed into the outer corona in form of the plasma blobs. \\

\section{Discussion \& Conclusions}
A loop-like quiescent prominence has been erupted on 18$^{th}$ November 2014 at 07:30 UT. Before the eruption, the prominence is remained into the stable state for more than three days and it's one leg is covered with the one-fourth part of the solar disk. In the present paper, we observe the evolution of multi-mode magnetic Rayleigh-Taylor unstable plumes (P1 \& P2) and their formation, which passes through the overlying prominence. Hybrid KH-RT instability at the prominence-cavity interface is also observed in the later phases.\\
In the upper part of the loop-like prominence, small-scale cavities are appeared and grow within the overlying prominence. After some vertical and horizontal growth of the cavity, some small-scale linear perturbations are developed at the boundary of the cavity (Figs.~1-2). A perturbation becomes nonlinear after 05:15 UT and formed a dark and hot plume (P1), which passes through the overlying prominence with an upflow speed of $\sim$35 km s$^{-1}$. The Differential Emission Measure (DEM) diagnostics show that the plume (dark upflow) is associated with the hotter and less denser coronal plasma as compared to the surrounding prominence. The higher density prominence region lies above the low density dark cavity region, and they are supported against the gravity. Additional perturbations continuously grow at the boundary of the cavity. Another perturbation is developed nonlinearly and converted into a bigger MRT unstable plume (P2). We estimate the observational growth rate of the MRT unstable plumes (P1) and (P2), which is 1.32$\pm$0.29$\times$ 10$^{-3}$ and 1.48$\pm$0.29$\times$ 10$^{-3}$ s$^{-1}$ respectively. The linear stability theory of the MRT instability has been used to estimate the theoretical growth rate of an MRT instability, which is found to be 1.95$\times$ 10$^{-3}$ s$^{-1}$. It is in good agreement with the observational growth rate. The second plume (P2) is converted into the mushroom cap-like structures in its final stage of the evolution, which is characterized by Kelvin-Helmholtz unstable single plume. The MRT unstable plume collapses by the secondary Kelvin-Helmholtz instability in which the two vortex-like structures and rolled plasma structures are evolved in the plasma downfall at the boundary of the prominence-cavity (Fig.~3). The secondary Kelvin-Helmholtz instability shows the nonlinear formulation of MRT instability. Therefore, we observe the full evolution of MRT unstable plumes and their linear and nonlinear dynamics.\\
\begin{figure*}

{\caption: MRT unstable prominence bubbles are tracked at different height at the prominence-cavity interface. The observed bubbles propagate with different velocity and trying to merge within themselves. The collapse phase of MRT unstable bubbles has been observed in the lower panel.}
\begin{center}
\includegraphics[scale=0.8,angle=90,width=18.0cm,height=8.0cm,keepaspectratio]{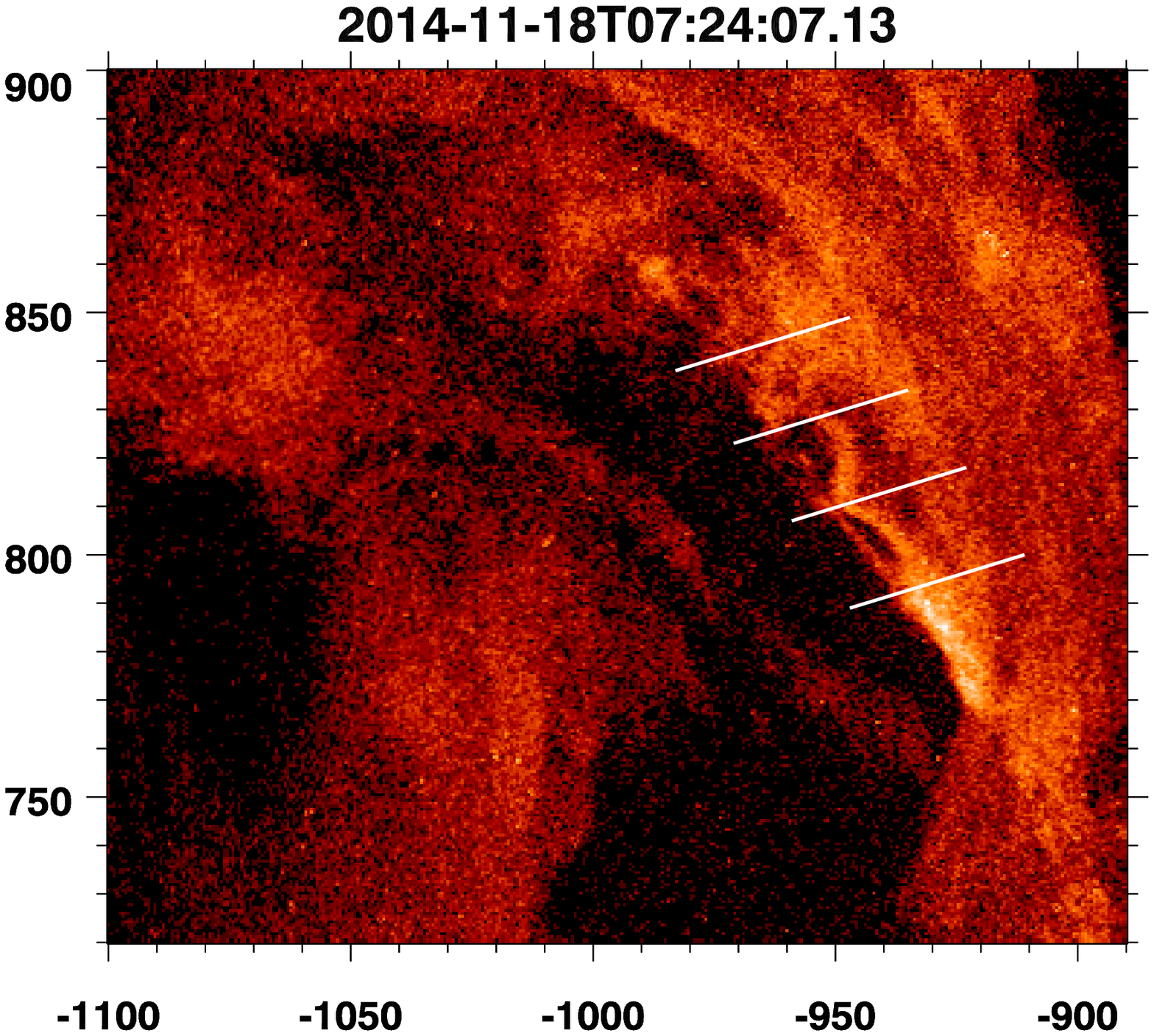}
\vspace{-1cm}
\end{center}
\end{figure*}
\begin{figure*}
\begin{center}
\includegraphics[scale=0.8,angle=0,width=18.0cm,height=10.0cm,keepaspectratio]{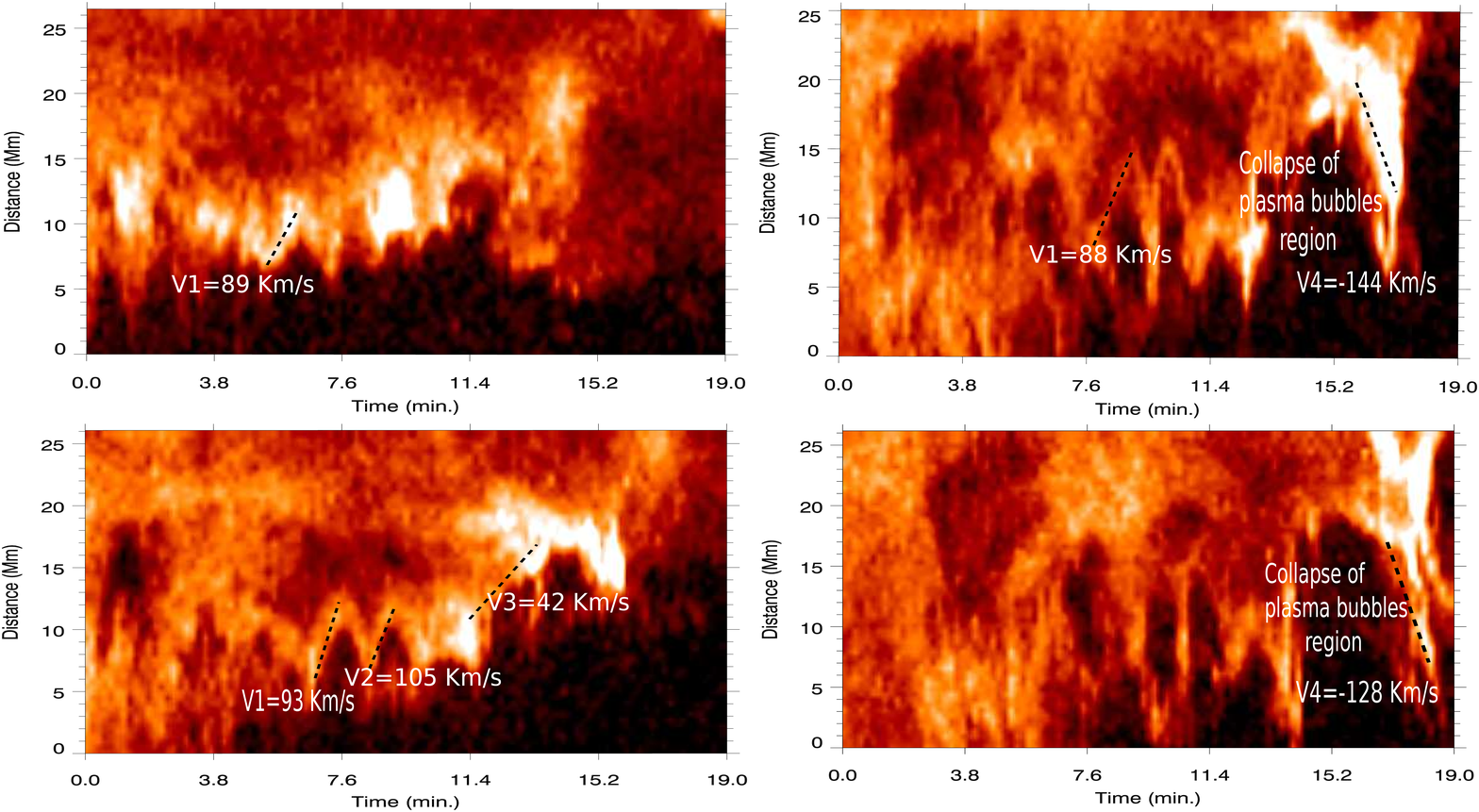}
\end{center}
\end{figure*}
 After the collapse of the MRT unstable plumes (P1 \& P2), the northward segment of the loop-like prominence consists of a bright plasma thread, which is evident at 06:43 UT in the overlying prominence (Fig.~4). The plasma thread is associated with the tangled field and Rayleigh-Taylor instability, which may be responsible for the shearing motion at the boundary of the prominence-cavity. During the evolution of the plasma thread, plasma bubbles, shear motion and plasma downfall are observed at the prominence-cavity interface (Figs,~6, 7, animation2.mpg, animation3.mog). The plasma bubbles are trapped within the eruption site and overlying prominence. The plasma bubbles may feel the depression due to the overlying prominence and may be responsible for the shear motion at the same boundary. Falling plasma at the interface may increase the shear motion which leads to the Kelvin-Helmholtz instability (Figs,~5-7, animation1, animation2). Later, the downfall of the plasma starts in the middle part of the plasma thread, which fixates the cool and dense plasma materials on the top of the thread at the interface. Due to the density inversion at the interface, MRT unstable plasma bubble are observed at the boundary of the prominence-cavity. The theoretical and observational merging time for these plasma bubbles are observed and they are consistent with each other. The shear flow at the prominence-cavity boundary may lead to the formation of Kelvin-Helmholtz unstable vortex structure and MRT unstable plasma bubbles at the same interface. Due to the hybrid KH-RT instability at the interface, the tangled magnetic field may try to balance the shear motion at the interface, which is sufficient for the collapse of an MRT unstable plasma bubbles.\\

In this paper, we have discussed the dynamics of  MRT unstable multimode plumes, which grew linearly and then converted into the single mode plume that indicate the nonlinear phase of the MRT instability of the plume development into the loop-like prominence. The similar dynamics of the MRT unstable plumes were reported previously into the hedgerow quiescent prominence only (e.g., Berger et al. 2008, 2010, 2011; Ryutova et al. 2010; Hillier et al. 2011a, 2012a, 2012b, 2018). Ryutova et al. (2010) have reported the similar order of the observational and theoretical growth rate in the quiescent prominence as we observe in the case of an eruptive loop-like prominence. Later, in addition, we observed a prominence thread which may contain tangled field and may be the cause of the shearing at the prominence-cavity interface. The thread-like structure is Rayleigh-Taylor unstable and may create the shearing (van Ballegooijen \& Cranmer 2010). Berger et al. (2017) have analyzed the boundary of the hedgerow prominence-cavity boundary and reported the hybrid KH-RT instability for the first time. The interface at which the thread passes became unstable and shows the characteristic of the hybrid KH-RT instability in that case. We observe the similar dynamics of hybrid KH-RT instability at the prominence-cavity interface now in the case of loop-like eruptive prominence.
\\
The present work, to the best of our knowledge, provides the first proof of the development of MRT unstable plumes and hybrid KH-RT instability together in the loop-like eruptive prominence. We find that these instability features develope on the different spatio-temporal scales and in the different region of the prominence in a sequential manner. Therefore, they provide over-all scenario of the evolution of linear MRT instability, its conversion to the non-linear phase, formation of the shear flows and vortices, and finally the generation of hybrid KH-RT instability phase. 
\section{Acknowledgements}
We acknowledge the constructive comments of the referee that improved the manuscript. AKS and SKM acknowledge the DST-SERB (YSS/2015/000621) project. AKS acknowledges the UKIERI Research Grant for the support of his research. Authors Acknowledge the use of Cheung et al. (2015) method developed by Mark Cheung to understand the thermal structure of the inner corona by observing the Differential Emission Measure (DEM). They also acknowledge the SDO/AIA observational data.

\end{document}